# *SMCHR: Satisfiability Modulo Constraint Handling Rules*

GREGORY J. DUCK

*School of Computing, National University of Singapore*
(*e-mail:* `gregory@comp.nus.edu.sg`)

### Abstract

*Constraint Handling Rules* (CHRs) are a high-level rule-based programming language for specification and implementation of constraint solvers. CHR manipulates a global store representing a flat conjunction of constraints. By default, CHR does not support goals with a more complex propositional structure including disjunction, negation, etc., or CHR relies on the host system to provide such features. In this paper we introduce *Satisfiability Modulo Constraint Handling Rules* (SMCHR): a tight integration of CHR with a modern *Boolean Satisfiability* (SAT) solver for quantifier-free formulae with an arbitrary propositional structure. SMCHR is essentially a *Satisfiability Modulo Theories* (SMT) solver where the theory $T$ is implemented in CHR. The execution algorithm of SMCHR is based on *lazy clause generation*, where a new clause for the SAT solver is generated whenever a rule is applied. We shall also explore the practical aspects of building an SMCHR system, including extending a "built-in" constraint solver supporting equality with unification and justifications.

## 1 Introduction

*Constraint Handling Rules* (CHRs) (Frühwirth 1998) are a high-level rule-based programming language for the specification and implementation of constraint solvers. CHR has two main types of rules: *simplification rules* ($H \Longleftrightarrow B$) rewrite constraints $H$ to $B$, and *propagation rules* ($H \Longrightarrow B$) *propagate* (i.e. add) constraints $B$ for every $H$. Constraint solvers are specified by sets of rules.

*Example 1* (*Bounds Propagation Solver*)
A *bounds propagation solver* propagates constraints of the form $X \geq L$ and $X \leq U$ for constants $L$ (lower bound) and $U$ (upper bound). We can specify how bounds are propagated through an addition $plus(X, Y, Z)$ constraint (i.e. $X = Y + Z$) via the following rules

$$plus(X, Y, Z) \wedge Y \geq L_Y \wedge Z \geq L_Z \Longrightarrow X \geq (L_Y + L_Z)$$
$$plus(X, Y, Z) \wedge Y \leq U_Y \wedge Z \leq U_Z \Longrightarrow X \leq (U_Y + U_Z)$$

Thus given the constraints $plus(A, B, C)$, $B \geq 3$, $B \leq 10$, $C \geq 4$, and $C \leq 6$, the rules will *propagate* $A \geq 7$ and $A \leq 16$. We can similarly write rules to propagate bounds in other directions. □

The execution algorithm for CHR is based on constraint *rewriting* and *propagation* over a global *store* of constraints. CHR solvers are *incremental*: when a new constraint $c$ is asserted, we check $c$ and the store against the rules to find a match. If there is a match, we apply the rule, possibly generating new constraints. Otherwise $c$ is inserted into the global store. Operational semantics and execution algorithms for CHR have been extensively studied (Duck et al. 2004)(Sneyers et al. 2010).

The global store represents a flat conjunction of constraints. CHR does not, by default, support goals that are formulae with a more complex propositional structure, e.g. with disjunction, negation, etc. Solving CHR constraints in other propositional contexts typically relies on some external machinery. For example, Prolog CHR implementations such as K.U.Leuven CHR system (Schrijvers and Demoen 2004) use Prolog's default backtracking search to handle disjunction.

In this paper we take a different approach: we extend CHR with a *Boolean Satisfiability* (SAT) solver to form *Satisfiability Modulo Constraint Handling Rules* (SMCHR). The basic idea is that we specify constraint solvers using CHR in the usual way, such as the rules in Example 1. SMCHR *goals* are then quantifier-free formulae of CHR constraints over any arbitrary propositional context.

*Example 2* (*SMCHR Goal*)
For example, the following SMCHR goal encodes the classic $n$-queens problem for the instance $n = 2$.

$$(Q_1 = 1 \lor Q_1 = 2) \land (Q_2 = 1 \lor Q_2 = 2) \land$$
$$\neg(Q_1 = Q_2) \land \neg(Q_1 = Q_2 + 1) \land \neg(Q_2 = Q_1 + 1)$$

This goal can be evaluated using an extended version of the bounds propagation solver from Example 1. For $n = 2$ the goal is unsatisfiable. □

Furthermore by integrating CHR with a modern implementation of SAT, we inherit all the advantages of no-good clause learning, non-chronological back-jumping, unit propagation, etc.

SMCHR is essentially a *Satisfiability Modulo Theory* (SMT) solver, where the *theory* solver T is implemented with CHR. SMT solvers have applications such as program verification, program analysis, model checking, theorem proving, constraint programming, etc. (Moura and Bjørner 2011). Most SMT solvers support a fixed set of first-order theories, such as linear arithmetic over the reals, arrays, uninterpreted functions, etc. SMCHR is much more flexible, as we can support any theory implementable in CHR.

Integrating CHR with a SAT solver presents several challenges. The first is the communication between the CHR and SAT solver engines. For this we use a generalization of *lazy clause generation* (Ohrimenko et al. 2009), a technique previously used to integrate SAT and finite domain solvers with impressive results. Another challenge is CHR's ability to "extend" existing *built-in* solvers, usually a solver for equality. For example, given the rule $(neq(X, X) \iff \textit{false})$ and the constraint $neq(A, B)$, the CHR solver must *ask* the built-in solver whether constraint $A = B$ holds, and if so, apply the rule. This is particularly challenging in the context of SMCHR as we wish to support both *unification*, efficient *variable indexing*, and *justifications* for clause generation. A "justification" is a description of *why* the given constraint holds.

We make the following contributions in this paper.
- Section 3 defines the SMCHR language and logical/operational semantics;
- Section 4 describes DPLL(CHR): the SMCHR execution algorithm based on DPLL and (lazy) clause generation;
- Section 5 describes an SMCHR runtime system based on a new variable representation supporting variable indexing;
- Section 6 describes a "built-in" reified equality solver supporting unification and justification; and
- Section 7 we run experiments to evaluate our SMCHR runtime system against an existing CHR implementation.

## 2 Preliminaries

*Constraint Handling Rules* (CHR) (Frühwirth 1998) have three types of rules

$$H \Longleftrightarrow B \qquad (simplification)$$
$$H \Longrightarrow B \qquad (propagation)$$
$$H_1 \setminus H_2 \Longleftrightarrow B \qquad (simpagation)$$

where the *head* $H$, $H_1$, $H_2$, and *body* $B$ are conjunctions of constraints. Simplification rules *replace* constraints matching $H$ with $B$. Propagations rules *add* constraints $B$ whenever constraints matching $H$ are found. Simpagation rules are a hybrid between simplification and propagation rules, where matching $H_2$ is replaced by $B$ whenever matching $H_1$ is found. The body $B$ may also contain *built-in* constraints such as *true* and equality $x = y$. Extended CHR also includes *guards* for checking built-in constraints during rule matching.

The *logical semantics* $[\![R]\!]$ of a given rule $R$ is defined as follows.

$$[\![H \Longleftrightarrow B]\!] = \forall (H \leftrightarrow B)$$
$$[\![H \Longrightarrow B]\!] = \forall (H \rightarrow B)$$
$$[\![H_1 \setminus H_2 \Longleftrightarrow B]\!] = \forall (H_1 \wedge H_2 \leftrightarrow H_1 \wedge B)$$

where $\forall F$ represents the universal closure of $F$. Here we assume $vars(B) \subseteq vars(H)$.

## 3 The SMCHR System

### 3.1 The SMCHR Language

The SMCHR language is the same as standard CHR with some minor differences. Constraint solvers are specified in terms of rules in the usual way. However, unlike CHR, SMCHR rules can handle both constraints and their *negations*.

*Example 3* (*Less-than Solver in SMCHR*)
Consider the following CHR solver that defines a "less-than" constraint $lt(X, Y)$.

$$lt(X, X) \implies \textit{false} \qquad \text{(reflexivity)}$$
$$lt(X, Y) \land lt(Y, X) \implies \textit{false} \qquad \text{(antisymmetry)}$$
$$lt(X, Y) \land lt(Y, Z) \implies lt(X, Z) \qquad \text{(transitivity)}$$

The rules (reflexivity), (antisymmetry), and (transitivity) respectively encode the properties $(\forall X : \neg X < X)$, $(\forall X, Y : \neg X < Y \lor \neg Y < X)$, and $(\forall X, Y, Z : X < Y \land Y < Z \to X < Z)$. With SMCHR we can also write rules that propagate negated constraints, e.g.

$$lt(X, Y) \implies \neg lt(Y, X) \qquad \text{(antisymmetry (2))}$$

Likewise we can write rules that match negated constraints.

$$\neg lt(X, Z) \land lt(X, Y) \implies \neg lt(Y, Z) \qquad \text{(transitivity (2))}$$
$$\neg lt(X, Z) \land lt(Y, Z) \implies \neg lt(X, Y) \qquad \text{(transitivity (3))} \quad \square$$

We extend the *logical semantics* of CHR in the obvious way allowing for negation. Under the logical semantics the rules (antisymmetry) and (antisymmetry (2)) are equivalent. Likewise rules (transitivity), (transitivity (2)), and (transitivity (3)) are also equivalent. However these rules are not *operationally* equivalent.

Adding negation is not a significant extension of CHR. In fact, negation can already be encoded in standard CHR, e.g. by introducing a new constraint symbol:

$$\textit{not\_lt}(X, Z) \land lt(X, Y) \implies \textit{not\_lt}(Y, Z)$$

The difference is that with SMCHR it is understood that $\neg c(\bar{x})$ is the negation of $c(\bar{x})$ and vice versa. We also express CHR rules in *reified-notation*. This expresses positive literals $c(\bar{x})$ as $(\textit{true} \leftrightarrow c(\bar{x}))$, and negative literals $\neg c(\bar{x})$ as $(\textit{false} \leftrightarrow c(\bar{x}))$. For example, rule (transitivity (2)) can be written as

$$(\textit{false} \leftrightarrow lt(X, Z)) \land (\textit{true} \leftrightarrow lt(X, Y)) \implies (\textit{false} \leftrightarrow lt(Y, Z))$$

Reified-notation will be used later for matching reified constraints.

Other key differences between CHR and SMCHR include:
- *Range-Restricted*: We assume all SMCHR rules are *range restricted*, i.e. for rule head $H$ and body $B$ we have that $vars(B) \subseteq vars(H)$.
- *Set-Semantics and Negation*: SMCHR assumes at most one copy of a constraint can appear in the store at once. This is equivalent to assuming the following rules are "built-in" for each constraint symbol $c$:

$$c(\bar{x}) \setminus c(\bar{x}) \iff \textit{true} \qquad \text{and} \qquad c(\bar{x}) \land \neg c(\bar{x}) \implies \textit{false}$$

- *Head-Connectiveness*: For all rules $R$ with head $H$, for all $h \in H$ let $H' = H - \{h\}$, then if $H' \neq \emptyset$ we require $vars(h) \cap vars(H') \neq \emptyset$. That is, for all multi-headed rules, every head constraint must share at least one variable with another head constraint.

Like SMT, SMCHR operates on *quantifier-free* formulae. *Range-Restricted* CHR programs ensure that no new (existentially quantified) variables are ever introduced by rule application. SMCHR with quantification is a possible direction for future work. *Set-Semantics* allows each constraint $c(\bar{x})$ to be associated with exactly one propositional variable $b$. This simplifies the design of the SMCHR system. *Head-Connectiveness* ensures that constraints can be matched against rules using *variable indexing* techniques. This will be discussed further in Section 5. The *Head-Connectiveness* requirement is specific to our SMCHR implementation, rather than a requirement in general.

In our experience, most CHR programs that implement constraint solvers satisfy the above conditions.

### 3.2 The SMCHR Operational Semantics: $\omega_s$

The SMCHR operational semantics are an extension of the *theoretical operational semantics* $\omega_t$ of CHR (Duck et al. 2004) (Frühwirth 1998). An *execution state* $\langle G, S, B, T \rangle$ is a 4-tuple consisting of a *goal* $G$, a constraint *store* $S$, a *built-in* constraint store $B$, and a *propagation history* $T$. A *reified constraint* is of the form $(b \leftrightarrow c)$ where $b$ is a propositional variable. Both $G$ and $S$ are sets of reified constraints, $B$ is a conjunction of *built-in* constraints, and $T$ is a set of tuples of the form $(r, b_1, .., b_n)$ where $r$ is a *rule identifier*, and $b_1, .., b_n$ are propositional variables. We assume the existence of a built-in solver $\mathcal{D}$ that supports Boolean and equality constraints.

Given an initial quantifier-free formula $G_0$, let $\mathsf{normalize}(G_0) = G_0^B \wedge G_0^R$ be a *normalized* formula such that (1) $G_0^B$ is a *propositional* formula, (2) $G_0^R$ is a conjunction of reified constraints of the form $(b_1 \leftrightarrow c_1) \wedge .. \wedge (b_n \leftrightarrow c_n)$, where each propositional variable $b_i$ is unique, and (3) $G_0$ is *equisatisfiable* to $G_0^B \wedge G_0^R$, i.e. $G_0$ is satisfiable iff $G_0^B \wedge G_0^R$ is satisfiable. Then the *initial state* for $G_0$ is $\langle G_0^R, \emptyset, G_0^B, \emptyset \rangle$. An exact definition of $\mathsf{normalize}$ is left to the implementation, provided the above conditions are preserved.

We define $\mathsf{isSet}(B, b)$ to hold iff $\mathcal{D} \models B \rightarrow b$ or $\mathcal{D} \models B \rightarrow \neg b$. Our SMCHR operational semantics $\omega_s$ introduces a **Decide** transition that *sets* a propositional variable to either *true* or *false*. The $\omega_s$ semantics are defined as follows:

1. **Decide**: $\langle G, S, B, T \rangle \rightarrowtail \langle G, S, B \wedge (b \leftrightarrow t), T \rangle$ where $b \in vars(B)$ is a propositional variable, $t \in \{true, false\}$ is a propositional constant, and $b$ is not already *set*, i.e. $\neg\mathsf{isSet}(B, b)$.
2. **Solve**: $\langle \{(b \leftrightarrow c)\} \uplus G, S, B, T \rangle \rightarrowtail \langle G, S, (b \leftrightarrow c) \wedge B, T \rangle$ where $c$ is a *built-in* constraint.
3. **Introduce**: $\langle \{(b \leftrightarrow c)\} \uplus G, S, B, T \rangle \rightarrowtail \langle G, \{(b \leftrightarrow c)\} \cup S, B, T \rangle$ where $c$ is a CHR constraint.
4. **Apply**: $\langle G, C_1 \uplus C_2 \uplus S, B, T \rangle \rightarrowtail \langle E \uplus G, C_1 \uplus S, M \wedge B, \{t\} \cup T \rangle$ where there exists a reified-notation rule $R = (r @ H_1 \setminus H_2 \Longleftrightarrow D)$ and a matching substitution $\theta$ such that $\theta.H_1 = C_1$, $\theta.H_2 = C_2$, and $(\mathcal{D} \models B \rightarrow \exists_{vars(H_1, H_2)} \theta)$; and for $D = (t_1 \leftrightarrow d_1) \wedge .. \wedge (t_m \leftrightarrow d_m)$ we have that $E = \{(b'_1 \leftrightarrow \theta.d_1), .., (b'_m \leftrightarrow$ $\theta.d_m)\}$ and $M = (b'_1 \leftrightarrow t_1) \wedge .. \wedge (b'_m \leftrightarrow t_m)$ for fresh propositional variables $b'_1, .., b'_m$. Finally $t = (r, b_1, .., b_n)$ where $C_1 \uplus C_2 \equiv \{(b_1 \leftrightarrow \_), .., (b_n \leftrightarrow \_)\}$, and $t \notin T$.

A *failed* state occurs when the built-in store $B$ is unsatisfiable, i.e. $\mathcal{D} \models \forall \neg B$. A *final* state $\sigma_F$ is either a failed state, or a state where no $\omega_s$ transition is applicable.

Given a goal $G$ with initial state $\sigma_I$, SMCHR *answers* UNSAT iff (1) for all derivations $\sigma_I \rightarrowtail^* \sigma_F$ where $\sigma_F$ is a final state then $\sigma_F$ is also a failed state, and (2) there are no non-terminating derivations $\sigma_I \rightarrowtail^* ...$ In other words: all possible derivations for $\sigma_I$ result in failure. Otherwise the answer is UNKNOWN and a non-failed final state $\sigma_F$ is the *result*, or we simply fail to terminate.

*Theorem 1 (Soundness)*

If the answer for goal $G$ is UNSAT, then $G$ is unsatisfiable, i.e. $[\![P]\!], \mathcal{D} \models \forall \neg G$.

*Proof*

(Sketch). By contradiction. Assume there exists a *satisfiable* goal $G$ with answer UNSAT. Since $G$ is satisfiable then $\mathsf{normalize}(G) = G_0^B \wedge G_0^R$ is also satisfiable. There must therefore exist a substitution $\theta_B$ over the propositional variables $vars(G_0^B)$ such that $\theta_B.G_0^B = true$ and $\theta_B.G_0^R$ is satisfiable. Consider the derivation

$$\sigma_I = \langle G_0^R, \emptyset, G_0^B, \emptyset \rangle \rightarrowtail^*_{\omega_s} \langle G_0^R, \emptyset, \theta_B \wedge G_0^B, \emptyset \rangle = \sigma'_I$$

comprised of only **Decide** transitions. Next consider a derivation $\sigma'_I \rightarrowtail^* \sigma_F$ from $\sigma'_I$ to a *failed* state $\sigma_F$. Such a derivation must exist, otherwise the answer for $G$ cannot be UNSAT. Since the propositional variables $vars(G_0^B)$ are already set, the derivation $\sigma'_I \rightarrowtail^* \sigma_F$ cannot contain a **Decide** transition.

The remainder of the proof is to show that each of the **Solve**, **Introduce**, and **Apply** transitions preserve satisfiability, as with standard CHR. Therefore if $\sigma'_I$ is satisfiable, then so is $\sigma_F$, and therefore $\sigma_F$ cannot be a failed state. This is a contradiction, and therefore the answer for a satisfiable $G$ cannot be UNSAT. □

The interpretation of UNKNOWN is merely the inability to prove unsatisfiability. An *incomplete* solver may answer UNKNOWN for unsatisfiable goals, i.e. the converse of Theorem 1 does not hold in general.

*Example 4 (Incomplete Solver)*

For example, consider the CHR program $P$

$$p \iff q \qquad p \implies false$$

and the goal $p$. The goal $p$ is unsatisfiable since $[\![P]\!], \mathcal{D} \models \neg p$, yet the *answer* for $p$ is UNKNOWN because of the non-failed derivation $\langle \{b \leftrightarrow p\}, \emptyset, b, \emptyset \rangle \rightarrowtail^* \langle \emptyset, \{b' \leftrightarrow q\}, b \wedge b', \emptyset \rangle$. □

A *complete* solver has the property that UNKNOWN = SAT. The result $\sigma_F$ may be mapped to a *solution* for $G$ depending on the definition of normalize.

```
dpllCHR(S)
    while ∃b ∈ S.Vars : ¬isSet(b)
        l = selectLiteral(S)
        S := setLiteral(S, l)
        S.Level := S.Level + 1
        S := propagate(S)
        if ∅ ∈ S.Clauses
            S = backtrack(S)
            if S.Level ≡ 0
                return UNSAT
    return UNKNOWN(S)

propagate(S)
    S := unitPropagate(S)
    let S' := S
    S := chrPropagate(S)
    if S' ≡ S return S
    return propagate(S)

unitPropagate(S)
    if ∅ ∈ S.Clauses
        return S
    if ∃l' : {l'} ∈ S.Clauses
        S := setLiteral(S, l')
        return unitPropagate(S)
    return S

chrPropagate(S)
    let Cs = chrMatch(Rules, S)
    if Cs ≡ nil return S
    S.Clauses := S.Clauses ∪ Cs
    return S
```

Fig. 1: Pseudo-code for the DPLL(CHR) algorithm.

## 4 DPLL(CHR): The SMCHR Execution Algorithm

The abstract operational semantics does not specify how and when the **Decide** transition is applied. For this we use a variant of the *Davis-Putnam-Logemann-Loveland* (DPLL) decision procedure for propositional formulae combined with CHR solving, i.e. DPLL(CHR).

The pseudo-code is shown in Figure 1. Search is controlled by the dpllCHR($S$) routine. Here $S$ represents the SMCHR "state" which includes

- $S.Vars$ is the set of all propositional variables;
- $S.Clauses$ is the a set of all clauses;
- $S.Level$ is the decision level;

The state $S$ also includes $S.Trail$ for backtracking, and $S.Store$ which contains the reified CHR constraint $b \leftrightarrow C$ corresponding to each $b \in S.Vars$. The top-level loop selects a propositional literal $l$ with selectLiteral, sets $l$ to *true* with setLiteral, and propagates the change with propagate. Here selectLiteral typically uses some heuristic to determine the "best" literal to select, and setLiteral replaces all $C \in S.Clauses$ where $C = \{\neg l\} \cup C'$ with $C'$, i.e. resolution. Propagation may result in the empty clause $\emptyset \in S.Clauses$ indicating failure in which case we backtrack. The process continues until either all variables are set, and we have reached UNKNOWN, or we backtrack to level 0, indicating UNSAT.

Our pseudo-code is a simplification. Our actual SMCHR implementation uses a modern SAT solver with no-good learning, back-jumping, etc.

Propagation is handled by the propagate, unitPropagate, and chrPropagate routines. First unitPropagate exhaustively propagates all *unit clauses* $\{l'\} \in S.Clauses$ by setting $l'$. Next we call chrPropagate to "wake up" any reified CHR constraint $b \leftrightarrow C$ where $b$ has been set to *true* or *false*. Here we assume *Rules* are in reified notation. The chrMatch routine attempts to match (and apply) a CHR rule. This involves searching for a (renamed apart) rule $(H_1 \setminus H_2 \iff D) \in \mathsf{Rules}$ and sets of constraints $C_1, C_2 \subseteq S.Store$ such that (1) for $C_1 \uplus C_2 = \{b_1 \leftrightarrow c_1(\bar{x}_1), .., b_n \leftrightarrow c_n(\bar{x}_n)\}$ and $H_1 \uplus H_2 = \{t_1 \leftrightarrow c_1(\bar{y}_1), .., t_n \leftrightarrow c_n(\bar{y}_n)\}$ we have that $b_i$ has been *set* (via setLiteral) to the propositional constant $t_i$; and (2) there exists a *matching substitution* $\theta$ such that $\theta.c_i(\bar{y}_i) = c_i(\bar{x}_i)$ for each $i \in 1..n$. If such a match $\theta$ is found, then

1. We *delete* $C_2$ from the store: $S.Store := S.Store - C_2$
2. For the rule body $D = \{t'_1 \leftrightarrow c'_1(\bar{z}_1), .., t'_m \leftrightarrow c'_m(\bar{z}_m)\}$ (where each $t'_i$ is a propositional constant) we generate the set $B$ of constraints defined as follows:

$$B = \{d_1 \leftrightarrow \theta.c'_1(\bar{z}_1), .., d_m \leftrightarrow \theta.c'_m(\bar{z}_m)\}$$

where each $d_i$ is either the propositional variable corresponding to an existing[1] $(d_i \leftrightarrow \theta.c'_i(\bar{z}_i)) \in S.Store$, else $d_i$ is a fresh propositional variable. We set $S.Store := B \cup S.Store$ and $S.Vars := \{d_1, .., d_m\} \cup S.Vars$.

3. We define the set of *head literals* $L_H$ and *body literals* $L_B$ as follows:

$$L_H = \{l_1, .., l_n\} = \{b_i \mid i \in 1..n \wedge t_i\} \cup \{\neg b_i \mid i \in 1..n \wedge \neg t_i\}$$
$$L_B = \{l'_1, .., l'_m\} = \{d_i \mid i \in 1..m \wedge t'_i\} \cup \{\neg d_i \mid i \in 1..m \wedge \neg t'_i\}$$

Finally we *generate* the following set of clauses:

$$Cs = \{(\neg l_1 \vee .. \vee \neg l_n \vee l'_i) \mid i \in 1..m \wedge \neg\mathsf{isTrue}(S, l'_i)\}$$

Note that we do not generate *redundant* clauses, i.e. when $\mathsf{isTrue}(S, l_i)$ holds. If there is no rule where $Cs \neq \emptyset$ nor a constraint is deleted, then chrMatch returns *nil*.

The chrMatch routine is essentially standard CHR rule application except (1) we are matching reified constraints, and (2) we are *generating* clauses and constraints. Each generated clause is either a unit clause that implies $l'_i \leftrightarrow true$, or is the empty clause because $l'_i$ is already set to *false*, indicating failure. The propagate routine re-invokes unitPropagate if a rule was applied. Propagation continues until a fixed-point is reached or failure occurs.

Propositional variables are never set by the CHR solver directly. Instead variables are set indirectly via the generated clauses. Each generated clause can be used by the SAT solver for future no-good learning and unit propagation. Our clause generation scheme is essentially a generalization of (Ohrimenko et al. 2009) from finite domain propagation solvers to any solver specified in CHR.

*Example 5 (SMCHR Execution)*
Consider the rules (reflexivity), (antisymmetry), and (transitivity) from Example 3. Suppose the initial goal $G$ is

$$\big(lt(A,B) \vee lt(B,A)\big) \wedge lt(B,C) \wedge \neg lt(A,C)$$

---

[1] *Set-Semantics* ensures there is only one choice for $d_i$.

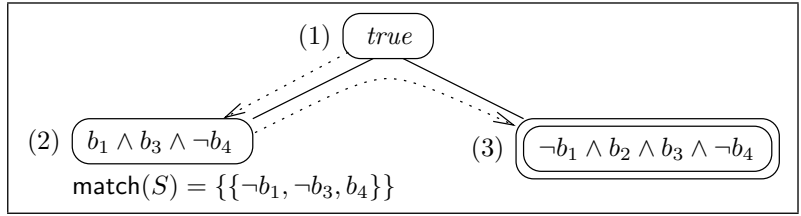

Fig. 2: An example SMCHR execution tree.

First we normalize $G$ into a propositional formula in CNF and reified CHR constraints as follows

$$\big[(b_1 \vee b_2) \wedge b_3 \wedge \neg b_4\big] \wedge b_1 \leftrightarrow lt(A,B) \wedge b_2 \leftrightarrow lt(B,A) \wedge b_3 \leftrightarrow lt(B,C) \wedge b_4 \leftrightarrow lt(A,C)$$

The initial state is therefore

$$S.Vars = \{b_1, b_2, b_3, b_4\} \qquad S.Clauses = \{\{b_1, b_2\}, \{b_3\}, \{\neg b_4\}\}$$
$$S.Store = \{(b_1 \leftrightarrow lt(A,B)), (b_2 \leftrightarrow lt(B,A)), (b_3 \leftrightarrow lt(B,C)), (b_4 \leftrightarrow lt(A,C))\}$$

A possible execution tree for $G$ is shown in Figure 2. Execution proceeds as follows:

(1) Assuming selectLiteral chooses literal $b_1$, after unitPropagate we have that $b_1 \wedge b_3 \wedge \neg b_4$ is set.
(2) Next chrPropagate is called. The rule $(lt(X,Y) \wedge lt(Y,Z) \Longrightarrow lt(X,Z))$ matches with $\theta = \{X/A, Y/B, Z/C\}$ and thus the clause $\neg b_1 \vee \neg b_3 \vee b_4$ is generated. This clause is empty (since $\neg b_4$ is already *true*) and therefore we fail and backtrack. The global set of clauses is now

$$S.Clauses = \{\{b_1, b_2\}, \{b_3\}, \{\neg b_4\}, \{\neg b_1, \neg b_3, b_4\}\}$$

A SAT with no-good learning would also generate $\{\neg b_1\}$.
(3) Next suppose selectLiteral selects the literal $\neg b_1$. After unitPropagate we have that $\neg b_1 \wedge b_2 \wedge b_3 \wedge \neg b_4$. No rule is applicable, and therefore we have reached a final state. The answer is therefore UNKNOWN with the result

$$\neg lt(A,B) \wedge lt(B,A) \wedge lt(B,C) \wedge \neg lt(A,C) \qquad \square$$

Another advantage of clause generation is that it completely subsumes the propagation history. In Example 5, the matching clause $\neg b_1 \vee \neg b_3 \vee b_4$ sets $b_4$ to *true* (i.e. isTrue$(S, b_4)$ holds) preventing the same rule from being applied once more on the same constraints. The rule will never be reapplied in other parts of the search tree where $b_1 \wedge b_3$ has been set to *true* and $b_4$ is unset. In such cases unitPropagate will set $b_4$ to *true* before chrPropagate is called, preventing the reapplication of the rule.

So far we have not considered CHR extending an existing "built-in" constraint solver. Later in Section 6 we will discuss how to extend SMCHR with a built-in solver supporting equality $x = y$.

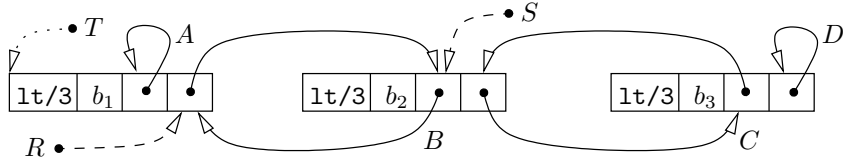

Fig. 3: Variable representation for the constraints from Example 6.

## 5 Implementation

We have implemented a SMCHR runtime system based on the DPLL(CHR) design from Section 4. Most of the design and implementation is standard and covered by existing literature on SAT (Een and Srensson 2003), SMT (Nieuwenhuis et al. 2006), CHR (Holzbaur 1999), and lazy clause generation (Ohrimenko et al. 2009). This section will focus on aspects of the SMCHR implementation that are novel.

### *5.1 Variable Indexing*

Prolog CHR implementations such as the K.U.Leuven CHR system (Schrijvers and Demoen 2004) implement *variable indexing* using *attributed variables* (Holzbaur 1999)(Holzbaur 1992). The basic idea is as follows: given a variable $X$, we attach an *attribute* to $X$ which contains all CHR constraints that mention $X$. This attribute can then be used to efficiently find partner constraints when matching CHR rules.

*Example 6* (*Attributed Variable Indexing*)
Suppose the CHR store contains the constraints $lt(A,B) \wedge lt(B,C) \wedge lt(C,D)$, then an attribute would be attached to $B$ via the call

$$\texttt{put\_attr}(B, \texttt{lt\_index}, [lt(A,B), lt(B,C)])$$

Here `lt_index` is the attribute's *name*, and the list $[lt(A,B), lt(B,C)]$ is the attribute's *value* containing all CHR constraints that mention variable $B$.

Consider the rule propagation rule $(lt(X,Y) \wedge lt(Y,Z) \implies lt(X,Z))$. Suppose we have matched $lt(A,B)$ with the occurrence $lt(X,Y)$, and we wish to find a *partner constraint* to match the occurrence $lt(Y,Z)$. With *attributed variable indexing*, we call $\texttt{get\_attr}(B, \texttt{lt\_index}, Ls)$ to retrieve the attributes value $Ls = [lt(A,B), lt(B,C)]$. We simply scan $Ls$ to find the match $lt(B,C)$, and apply the rule. □

Our SMCHR runtime system also uses variable indexing, but does not use attributed variables. Instead our variable indexing scheme is based on *PARMA-bindings* (Taylor 1996). PARMA-bindings represent a variable $X$ as a *pointer-cycle* through all constraints (or terms) that contain $X$. This is opposed to *pointer-chains* from a traditional WAM-style variable representation. A good comparison between WAM and PARMA variable representations can be found in (Demoen et al. 1999).

*Example 7* (*Term and Variable Representation*)
The representation of the (reified version of the) constraints from Example 6 is shown in Figure 3. Here we assume a reified constraint $b \leftrightarrow f(x_1, .., x_n)$ is represented by a $n$+1-arity term $f(b, x_1, .., x_n)$. The term itself is represented as a vector containing the functor/arity pair followed by the term's arguments $b, x_1, .., x_n$.

In Figure 3 the pointer-cycles are represented by the unbroken arrows. Symbols $b_1$, $b_2$, and $b_3$ are propositional variables handled by the SAT solver. The pointer-cycles for $B$ and $C$ are of length two because they appear in two constraints, whereas singleton variables $A$ and $D$ have a pointer-cycle length of one. □

A *variable reference* is a pointer to a cell in the pointer-cycle of a given variable. Given a reference $R$ to variable $X$, we define the following low-level operations:

1. var_next($R$) is a reference to the *next* cell in the pointer-cycle for $X$;
2. var_container($R$) is the term *containing* the cell pointed to by $R$; and
3. var_index($R$) is the argument index (from 0) of the cell pointed to by $R$.

*Example 8* (*Variable Operations*)
Consider Figure 3 once more. Here $R$ and $S$ are references to variable $B$, and $T$ is a reference to the term representing the first *lt* constraint. We see that

$$\text{var\_next}(R) = S \quad \text{var\_next}(S) = R \quad \text{var\_container}(R) = T$$
$$\text{var\_index}(R) = 2 \quad \text{var\_index}(S) = 1 \qquad \square$$

The var_next operation is simply pointer dereference, i.e. var_next($R$) ≡ *$R$. Our implementation of var_container relies on low-level garbage collector support. Namely, it relies on a garbage collector that can efficiently[2] map *interior* pointers, i.e. pointers to inside a term such as $R$, to *exterior* pointers, i.e. the pointer to the term itself such as $T$. We have implemented a garbage collector as part of our SMCHR implementation.[3] Finally the var_index operation is implemented in terms of var_container, and pointer arithmetic, i.e. var_index($R$) ≡ $R$ − var_container($R$) − 1.

Using these low-level operations, we can directly implement *variable indexing*.

*Example 9* (*Variable Indexing*)
Consider the following constraints for variable $B$.

$$b_1 \leftrightarrow lt(A, B) \wedge b_2 \leftrightarrow lt(D, B) \wedge b_3 \leftrightarrow lt(B, C)$$

A possible variable layout for these constraints is shown in Figure 4a. Suppose that $b_1 \wedge b_2 \wedge b_3$ is set to *true*, and that we have matched $b_1 \leftrightarrow lt(A, B)$ with occurrence *true* $\leftrightarrow lt(X, Y)$ from the rule $(lt(X, Y) \wedge lt(Y, Z) \implies lt(X, Z))$. To match the rule we must find a partner constraint of the form *true* $\leftrightarrow lt(B, \_)$ using $B$ as the variable index.

The pseudo-code for the matching routine in shown in Figure 4b. The algorithm traverses the pointer-cycle of $B$. It will stop when either a match is found, or if we complete a full loop without finding a match. Assuming the layout from Figure 4a, the match_lt routine will find the matching constraint $b_3 \leftrightarrow lt(B, C)$ in the third iteration of the loop. □

---

[2] In constant $O(1)$ time.
[3] Other garbage collectors also support the mapping of interior to exterior pointers, e.g. the Boehm Collector (Boehm and Weiser 1988) with the GC_base($ptr$) API call.

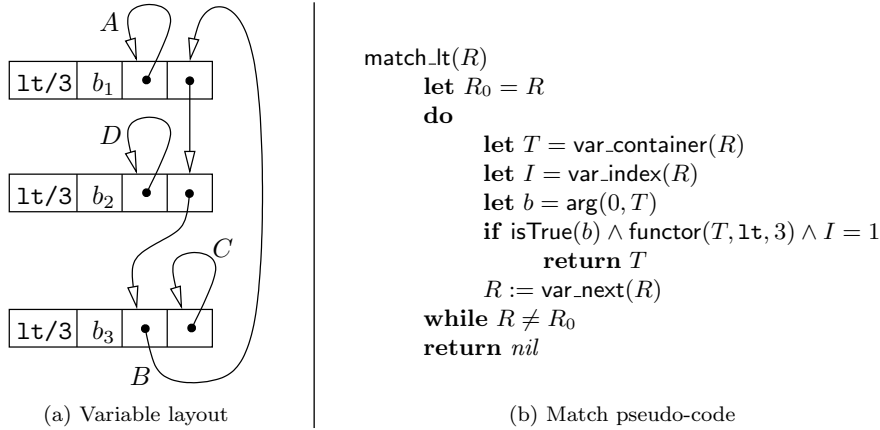

(a) Variable layout         (b) Match pseudo-code

Fig. 4: Matching a $lt(\_, B, \_)$ constraint.

Our variable representation is advantageous when it comes to discovering *justifications* for equality, which will be discussed in Section 6. The disadvantage is that our variable indexing scheme does not directly allow for variables nested inside other terms, e.g. $c(f(X,Y), Z)$. This is acceptable for solvers over "flat" domains such as integers. For other domains we may apply the *flattening transformation* as described in (Sarna-Starosta and Schrijvers 2008). Here we flatten constraints by introducing new constraint symbols, e.g. $c(f(X,Y), Z)$ becomes $cf(X, Y, Z)$.

### 5.2 Constraint Deletion

When a simplification or simpagation rule is applied, we must *delete* some of the constraints that matched the rule head. Say we wish to delete constraint $c(\bar{X})$, this can be achieved in one of two possible ways:

1. Remove $c(\bar{X})$ from the pointer-cycles of all $X \in \bar{X}$; or
2. Overwrite the functor $c$ with some "dummy" functor, say $d$.

The latter destructively updates $c(\bar{X})$ to $d(\bar{X})$ making it invisible to future matchings. The former is the more expensive operation, but keeps pointer-cycles shorter, which may be advantageous in the long run.

## 6 A Reified Equality Solver

Most CHR systems extend a "built-in" constraint solver that, at a minimum, supports equality $x = y$. Prolog CHR systems such as (Schrijvers and Demoen 2004) extend Prolog's standard unification =/2 over (attributed) variables and terms. For SMCHR we require a built-in solver that supports *reified equality constraints* of the form $b \leftrightarrow (X = Y)$, where $b$ is a propositional variable. For efficient rule matching using variable indexing, we wish to support variable *unification*. However, for clause generation, we must also support *justifications* that include equality constraints. In this section we describe how to implement such a solver using the variable representation described in Section 5.1.

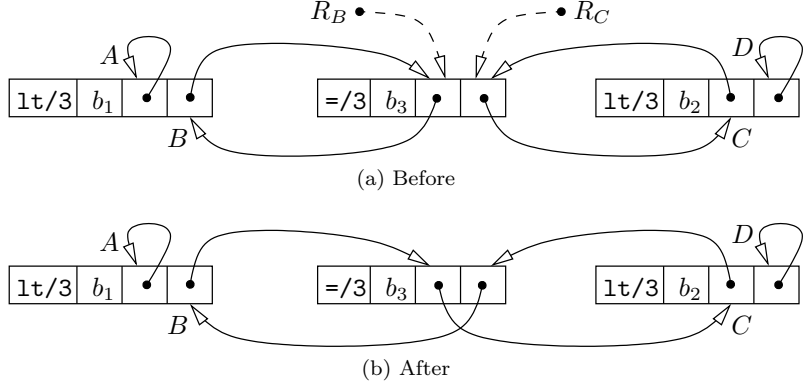

Fig. 5: Variable representation and unification.

### 6.1 Unification

Given an equality constraint $b \leftrightarrow (X = Y)$ where $b$ has been set to *true*, then we must *unify* the representations of variables $X$ and $Y$. This is handled the same way as with PARMA-bindings, i.e., by merging the pointer cycles for $X$ and $Y$ to form one larger cycle (Demoen et al. 1999). Specifically, let $T$ be the term representation of the equality constraint $b \leftrightarrow (X = Y)$; let $R_X$ and $R_Y$ be the variable *references* to $X$ and $Y$ in $T$; and let $R'_X = \mathsf{var\_next}(R_X)$ and $R'_Y = \mathsf{var\_next}(R_Y)$, then we set

$$*R_X := R'_Y \quad \text{and} \quad *R_Y := R'_X$$

Both updates are trailed and will be undone on backtracking.

*Example 10* (*Variable Unification*)
Consider the constraints:

$$b_1 \leftrightarrow lt(A,B) \wedge b_2 \leftrightarrow lt(C,D) \wedge b_3 \leftrightarrow (B = C)$$

The representation of these constraints is shown in Figure 5a. Here $R_B$ and $R_C$ are the references to $B$ and $C$ in the term representation of the equality constraint.

Suppose $b_3$ is set to *true*. Let $R'_B = \mathsf{var\_next}(R_B)$ and $R'_C = \mathsf{var\_next}(R_C)$, then the equality solver *unifies* $B$ with $C$ by setting $*R_B := R'_C$ and $*R_C := R'_B$. This merges the pointer-cycles to form one larger cycle, as shown in Figure 5b. □

Note that variable unification only the updates the equality term itself. All other constraints remain unchanged.

We avoid unifying two references to already-unified variables. Instead we delete the redundant equality constraint. In effect, our equality solver enforces the following CHR rule

$$X = Y \setminus X = Y \iff true$$

The built-in equality solver only handles variable/variable unification. Variable/value equality can be implemented via the rule

$$X = c \wedge X = d \implies false \quad \text{where} \quad c \neq d$$

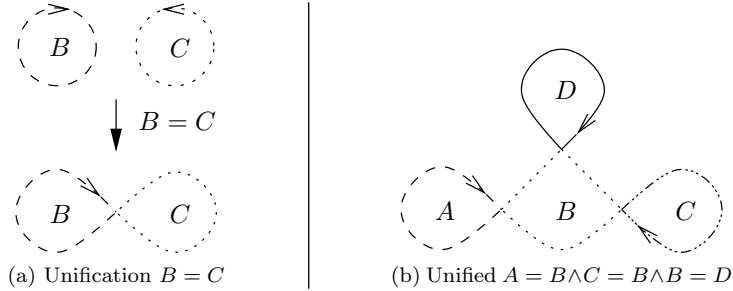

Fig. 6: Unification and justification of pointer-cycles.

### 6.2 Justification

Given two references to variables $x$ and $y$, we must determine whether $x$ and $y$ are equal, and if they are, generate a *justification* as to *why* $x$ and $y$ are equal.

*Example 11 (Equality Justification)*
Given the rule $(lt(X,Y) \wedge lt(Y,X) \implies \mathit{false})$ and the constraints

$$b_1 \leftrightarrow lt(A,B) \wedge b_2 \leftrightarrow lt(C,A) \wedge b_3 \leftrightarrow (B=D) \wedge b_4 \leftrightarrow (D=C) \wedge b_5 \leftrightarrow (A=E)$$

suppose $b_1 \wedge b_2 \wedge b_3 \wedge b_4 \wedge b_5$ is set to *true*. The rule matches since $\mathcal{D} \models B = D \wedge D = C \to B = C$ with the corresponding justification $b_3 \wedge b_4$. The negated justification is incorporated into the clause $\neg b_1 \vee \neg b_2 \vee \neg b_3 \vee \neg b_4$ that is generated when the rule is applied. □

The challenge is to combine justification with unification. In Example 11, variables $B$, $C$, and $D$ will already be unified as per Section 6.1, and thus share one combined pointer-cycle. For justifications, we must treat $B$, $C$, and $D$ as separate variables. We can however make the following observation:

*Observation 1*
The term representation of a reified equality constraint $b \leftrightarrow (X = Y)$ remains in the combined pointer-cycle after unification.

This is demonstrated in Figure 5. After the unification (in Figure 5b), the term representation of $b_3 \leftrightarrow (B = C)$ remains in the combined pointer-cycle. This term marks the point where the pointer-cycles for $B$ and $C$ were joined. With this information, we can *reconstruct* the original pointer-cycles before the unification.

To help describe the algorithm for computing justifications, we introduce an abstraction based on the following observation about the "shape" of the combined pointer-cycle in Figure 5b:

*Observation 2*
The combined pointer-cycle has a "twist" in it, i.e. where it overlaps with itself to form a horizontal "figure-8" pattern.

The *abstract* version the unification from Figure 5 is shown in Figure 6a. Variables are represented by abstract pointer-cycles. Unifying two pointer-cycles forms

```
ask_eq(X, Y)
    let X_0 = X, S = ε
    do
        if X ≡ Y return S
        let T = var_container(X)
        let b = arg(0, T)
        if isTrue(b) ∧ functor(T, =, 3)
            if peek(S) ≡ b
                S := pop(S)
            else
                S := push(S, b)
        X := var_next(X)
    while X ≠ X_0
    return unknown
```

(a) Ask-equals with justification

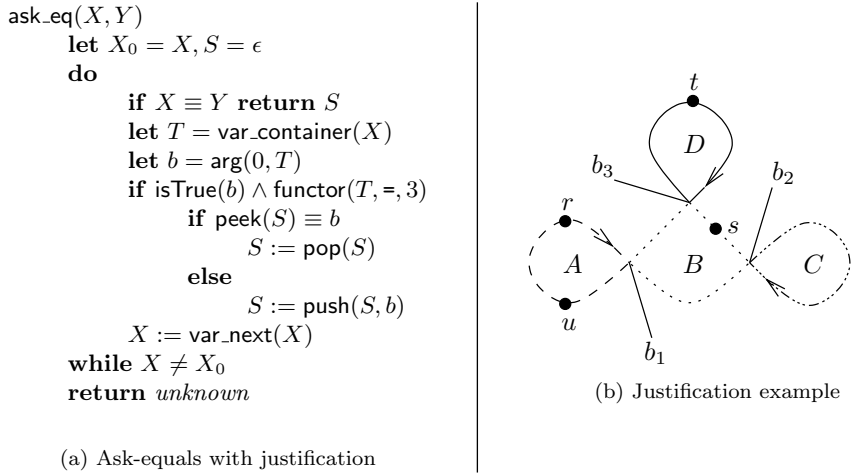

(b) Justification example

Fig. 7: Ask-equals with justification example.

a larger pointer-cycle with a twist. The twist is merely an abstraction of the equality term, and therefore denotes the transition between the original pointer-cycles from $B$ to $C$ and vice versa.

*Example 12 (Twists)*
Suppose $A$, $B$, $C$, and $D$ are "fresh" variables with no twists, i.e. are yet to be unified with any other variable. Suppose that

$$b_1 \leftrightarrow (A = B) \wedge b_2 \leftrightarrow (C = B) \wedge b_3 \leftrightarrow (B = D)$$

and $b_1 \wedge b_2 \wedge b_3$ is set to *true*. Variables $A$, $B$, $C$, and $D$ are unified to form an abstract pointer-cycle with three twists as shown in Figure 6b. □

Equality justifications can be straightforwardly generated by traversing pointer-cycles and tracking twists. Consider the rule ($neq(X, X) \Longrightarrow false$) that implements a disequality constraint. Given a CHR constraint $neq(A, B)$, the SMCHR engine *asks* the built-in equality solver whether variable references $A$ and $B$ are equal, and if so, the justification for the equality for clause generation. The pseudo-code for the *ask-equals* algorithm is shown in Figure 7a. The inputs are variable references $X$ and $Y$. The algorithm traverses the pointer-cycle starting from $X$ and maintains a stack $S$ of twists. Each twist is represented by the propositional variable $b$ from the corresponding non-redundant reified equality constraint $b \leftrightarrow (\_ = \_)$. As we traverse the pointer-cycle and we encounter a twist $b$, we either (1) *pop* $b$ from $S$ if the top of $S$ is $b$ (i.e. peek$(S) \equiv b$), or (2) *push* $b$ onto $S$ otherwise. If $Y$ is encountered, then $X = Y$ holds and the stack $S$ is returned. Otherwise *unknown* is returned. The justification for the equality is simply the resulting stack $S = [b_1, .., b_n]$ interpreted as the conjunction $b_1 \wedge .. \wedge b_n$.

*Example 13 (Justifications)*
Consider the abstract pointer-cycle diagram from Figure 7b and the variable refer-

16                                G. Duck| Bench. | Solver | SMCHR | | | CHR | |
|---|---|---|---|---|---|---|
| | | time | #clauses | #fails | time | #fails |
| `cycle(50)` | lt | 285 | 1177 | 1 | 238 | 1 |
| `cycle(100)` | " | 8003 | 4852 | 1 | 3515 | 1 |
| `cycle(50)` | leq | 501 | 1225 | 0 | 242 | 0 |
| `cycle(100)` | " | 12113 | 4950 | 0 | 3382 | 0 |
| `queens(12)` | bounds | 582 | 2834 | 388 | 847 | 2805 |
| `queens(14)` | " | 1750 | 5829 | 991 | 9029 | 24596 |
| `queens(16)` | " | 7072 | 14659 | 4119 | 65951 | 150660 |
| `queens(18)` | " | 22984 | 33159 | 12972 | 359170 | 701930 |
| `queens(20)` | " | 82200 | 84000 | 44548 | – | – |
| `subsets(15, 99)` | " | 57 | 7477 | 106 | 4301 | 30827 |
| `subsets(20, 99)` | " | 102 | 12687 | 156 | 79115 | 616666 |
| Geo. Mean. | – | 1296 | – | – | 447% | – |

Fig. 8: SMCHR vs. K. U. Leuven CHR (SWI) experimental results

ences $r, u$ to $A$, $s$ to $B$, and $t$ to $D$, then

| (query) | (sequence) | (justification) |
|---|---|---|
| $\mathsf{ask\_eq}(r, s)$ | $\mathsf{push}(b_1), \mathsf{push}(b_2), \mathsf{pop}(b_2)$ | $b_1$ |
| $\mathsf{ask\_eq}(r, t)$ | $\mathsf{push}(b_1), \mathsf{push}(b_2), \mathsf{pop}(b_2), \mathsf{push}(b_3)$ | $b_1 \wedge b_3$ |
| $\mathsf{ask\_eq}(r, u)$ | $\mathsf{push}(b_1), \mathsf{push}(b_2), \mathsf{pop}(b_2), \mathsf{push}(b_3), \mathsf{pop}(b_3), \mathsf{pop}(b_1)$ | $true$ |
| $\mathsf{ask\_eq}(u, r)$ | $\epsilon$ | $true$ |

Furthermore we see that $\mathsf{ask\_eq}(R, S) = \mathsf{ask\_eq}(S, R)$ for all $R, S$. For example, $\mathsf{ask\_eq}(u, r) = \mathsf{ask\_eq}(r, u) = true$, as expected. □

We can similarly adapt the *variable indexing* routines such as from Figure 4b to generate justifications for matches using the same basic idea.

## 7 Experiments

In this section we test the SMCHR runtime system presented in this paper. We are yet to integrate a CHR compiler, and all CHR solvers tested in this section were compiled manually. All timings are on Intel i5-2500K CPU clocked at 4Ghz and averaged over 10 runs. We compare against the K.U.Leuven CHR system (Schrijvers and Demoen 2004) running on SWI Prolog (Wielemaker et al. 2012) version 5.8.2., with debugging disabled and full CHR optimization enabled. The results are shown in Figure 8. Here, time is the time in milliseconds, #clauses is the number of generated clauses (for SMCHR), and #fails is the number of fails/backtracks. A dash indicates a time exceeding 10 minutes.

Benchmark `cycle(n)` is a cycle of CHR constraints $p(A_0, A_1) \wedge .. \wedge p(A_n, A_0)$ for $p \in \{lt, leq\}$. The lt solver answers *false*, whereas the leq unifies all the variables $A_i$ for $i \in 0..n$. Benchmark `queens(n)` finds a solution to the classic $n$-queens problem using a bounds propagation solver implemented in CHR (e.g. like Example 1) and a generalization of the encoding from Example 2. Benchmark `subsets(n, v)` is a variant of the sum-of-(multi-)sets problem, where the multi-set is $\{10, .., 10\}$ and is of size $n$, and $v$ is the target value. For $v = 99$ there are no solutions.

The experimental results show that SMCHR is slower for benchmarks that do not use search/disjunction, such as `cycle`($n$). This is because of overheads introduced by the SAT solver and clause generation. For benchmarks that use search, such as `queens`($n$) and `subsets`($n, v$), SMCHR is faster thanks to no-good learning. This benefit likely overwhelms any advantage gained from manual compilation.

## 8 Related Work

SMCHR is closely related to *Satisfiability Modulo Theories* (SMT) (Moura and Bjørner 2011). The basic design is the same: a SAT solver core sets and wakes *theory* constraints which are solved using a *theory solver* (Nieuwenhuis et al. 2006). In SMCHR, the theory solver is specified in CHR. As far as we are aware, most SMT implementations do not use the (lazy) clause generation (Ohrimenko et al. 2009) approach. Instead the theory solvers either do no propagation (i.e. merely test satisfiability) or the SAT core directly queries the theory solver for justifications on backtracking. The clause generation approach is far more flexible. It allows for theory propagation, and does not require any special algorithm to construct justifications on failure.

SMCHR is not the first CHR system to support back-jumping and no-good learning. CHR$^\vee$ (Wolf et al. 2008) is an extension of CHR that allows disjunction in the body of the rules, e.g. a rule for `indomain` for labelling might be:

$$\texttt{label}(X) \setminus \texttt{indomain}(X, [Y|Ys]) \iff X = Y \vee \texttt{indomain}(X, Ys)$$

In SMCHR we do not currently support disjunction in rules, only in the initial goal. Disjunction in rules is an obvious direction for future work. The operational semantics of CHR$^\vee$ is an extension of the the refined operational semantics (Duck et al. 2004), where *conflict driven back-jumping* is supported by explicitly annotating constraints with *justifications*. Unlike CHR$^\vee$, we do not build the machinery of no-good learning, back-jumping, etc., into the operational semantics. Rather these details are left to the SAT solver. It is not clear how the CHR$^\vee$ implementation supports justifications for equality.

## 9 Conclusions and Future Work

In this paper we introduced SMCHR, the natural merger of SMT, CHR, and (lazy) clause generation. We presented a new CHR runtime system based on a new variable representation that supports both unification, and importantly, justifications. Our experimental results confirm that no-good clause learning is very effective at search space pruning, resulting in large speed-ups for some benchmarks.

Our SMCHR runtime system has not yet been finely tuned and results from Section 7 can likely be improved. This is future work. We also intend to use SMCHR for applications such as automated program verification.